\documentclass[prb,twocolumn,final,showpacs,showkeys,nobibnotes,titlepage,twoside,10pt]{revtex4-1}
\usepackage{amsmath}
\usepackage{graphicx}
\usepackage{amsfonts}
\usepackage{amssymb}
\usepackage{subfigure}
\usepackage{float}
\usepackage{color}

\begin{document}

\title{Kinetics of fragmentation and dissociation of two-strand protein
filaments: coarse-grained simulations and experiments}

\author{A. Zaccone}
\email{az302@cam.ac.uk}
\affiliation{Department of Chemical Engineering and Biotechnology,  Cambridge
CB2 3RA, U.K}
\affiliation{Cavendish Laboratory, University of Cambridge, Cambridge CB3 0HE,
U.K.}
\author{I. Terentjev}
\affiliation{Granta Design, 62 Clifton Rd, Cambridge CB1 7EG, U.K.}
\author{T. W. Herling}
\author{T. P. J. Knowles}
\affiliation{Department of Chemistry, University of Cambridge,  Cambridge CB2
1EW, U.K.}
\author{A. Aleksandrova}
\author{E. M. Terentjev}
\affiliation{Cavendish Laboratory, University of Cambridge, Cambridge CB3 0HE,
U.K.}

\date{\today}

\begin{abstract}

While a significant body of investigations have been focused on the process of
protein self-assembly, much less is understood about the reverse process of a
filament breaking due to thermal motion into smaller fragments, or
depolymerization of subunits from the filament ends. Indirect evidence for
actin and amyloid filament fragmentation has been reported, although the
phenomenon has never been directly observed either experimentally or in
simulations. Here we report the direct observation of filament depolymerization
and breakup in a minimal, calibrated model of coarse-grained molecular
simulation. We quantify the orders of magnitude by which the depolymerization
rate from the filament ends $k_\mathrm{off}$ is larger than fragmentation rate
$k_{-}$ and establish the law $k_\mathrm{off}/k_- = \exp [( \varepsilon_\| -
\varepsilon_\bot) / k_\mathrm{B}T ] = \exp [0.5 \varepsilon / k_\mathrm{B}T ]$,
which accounts for the topology and energy of bonds holding the filament
together. This mechanism and the order-of-magnitude predictions are well
supported by direct experimental measurements of depolymerization of insulin
amyloid filaments.

\end{abstract}

\maketitle

\section{Introduction}
The strong directionality of non-covalent physical bonds between proteins
underlies their strong propensity to self-assemble into fibrils and filaments.
Protein filaments are ubiquitous in biology, appearing individually, in
bundles, or in randomly crosslinked networks. They facilitate the propulsion of
bacteria and extension in lamellipodia, they control the mechanical strength of
the cytoskeleton and the bending stiffness in axons, they allow positional
control of organelles and provide transport pathways all around the cell
\cite{1-Oosawa,2-Alberts}. In a different context, the self-assembly of
misfolded proteins into amyloid fibrils impairs physiological activity and is
associated with a number of organic dysfunctions
\cite{3-Chiti,4-Chandler,5-Irback}. In yet another context, filaments are
artificially or spontaneously assembled to achieve a specific function in the
material, such as directed conductivity, plasmonic resonances, or just the
mechanical strength in a fibre composite, all with important technological
applications. While a huge number of experimental and computational studies are
available on the mechanism of self-assembly of proteins into filaments, the
reverse process of filament breakup (fragmentation) remains poorly understood
and controversial
\cite{6-Wegner,7-Wegner,8-Pollard,9-Erickson,10-Kinosian,11-Sept,12-Pollard,13-Fushiwara,14-Kuhn,15-Fass}.

\subsection{The two cases of actin filaments and amyloid fribrils}
In the case of actin, the depolymerization rates were measured long ago by
Pollard~\cite{8-Pollard}, but no direct measurement of fragmentation is
available: fragmentation rates are typically deduced from kinetic modelling
assuming an equilibrium between fragmentation and filament-filament annealing
\cite{6-Wegner,7-Wegner,9-Erickson,11-Sept}.
For F-actin, both end-dissociation and fragmentation are influenced by chemical
factors: F-actin is ADP-bound at the pointed end and in the inner filament,
where fragmentation occurs, while it is mostly ATP-bound at the barbed end
(with a smaller end-dissociation rate), especially in the treadmilling and fast
polymerization regimes \cite{2-Alberts}. The ADP-complexation thus reduces the
binding energy somewhat compared to ATP, resulting to higher rates for thermal
escape of a subunit from the attractive potential well with its bonded
subunits. As shown experimentally \cite{Vavylonis}, and also by
theoretical modelling \cite{Kolom-Biophys}, growth rates and the critical
concentration at the barbed end are intimately related to the cap structure and
dynamics. However, these effects are less important for dissociation processes
where only ADP is left inside actin monomers.

In the case of amyloid filaments, which have a similar multi-stranded structure
but different protein-protein bonding, no data on end-dissociation is available
while the rates of fragmentation are similarly deduced from kinetic modelling
\cite{16-Knowles,17-Adamcik}. This gap in the current knowledge has important
consequences: in the case of actin, it is fragmentation which sets both the
lifetime of actin filaments (to ca. 500 s in vitro \cite{2-Alberts}) and their
plateau average length \cite{9-Erickson}, as well as secondary nucleation and
rapid growth of amyloid fibrils (self-catalytic activity). The physics of
fragmentation is crucial to explain the anomalously large length-diffusivities,
the microscopic mechanism of severing in vivo, and to engineer the mechanical
properties of artificial filaments for biotechnology.
Also, little is known about the rates of end-dissociation in amyloids, while
the fragmentation is estimated from the equilibrium kinetics balance, and the
customary assumption is to take both rates equal~\cite{16-Knowles,20-Lee},
which is an assumption very far from reality, as we will show below.

\subsection{Simplified self-assembly framework}
In a simplified framework that neglects the role of active oligomers and
nuclei, the filament growth can be summarised by the reversible reaction: $A_1
+ A_p  \rightleftharpoons A_{p+1}$, where the monomer subunit $A_1$ is added to
a filament of $p$ units. For the forward reaction, it is accepted that
association is dominated by the addition of a single subunit (elongation),
while annealing (joining of two fragments) is much slower, because of the
greater abundance of monomers with respect to active oligomers and the fast
decay of filament mobility with its length \cite{18-Kas}. The subunit
dissociation reaction from the end is known to be dominant in the case of actin
\cite{9-Erickson,15-Fass,19-Pollard}, although the much slower fragmentation
reaction is the one which controls the late-stage plateau in the growth.

Here we provide a minimal numerical model of break-up rates which attempts to
bring together the essential features of both actin and amyloid filaments.
Although more dynamic exchange processes (such as those described e.g. in
Ref.\cite{Carulla,Vestergaard}) are neglected in order to focus on the general
mechanisms, the model predictions for the rates of fragmentation and
dissociation are satisfactorily verified with new experiments for the case of
amyloid-like insulin fibrils. Although we aim to provide the missing breakup
mechanism for the specific case of amyloid-like filaments, our framework is
general. By combining it with detailed atomistic approaches it will be useful
for the control of biofilaments size in biotechnological applications (e.g.
biomaterials for regenerative tissues, biofilms, etc).

\section{Results and Discussion}

\subsection{Coarse-grained numerical model}
\begin{figure}[t]
\centering
\includegraphics[width=0.6\columnwidth]{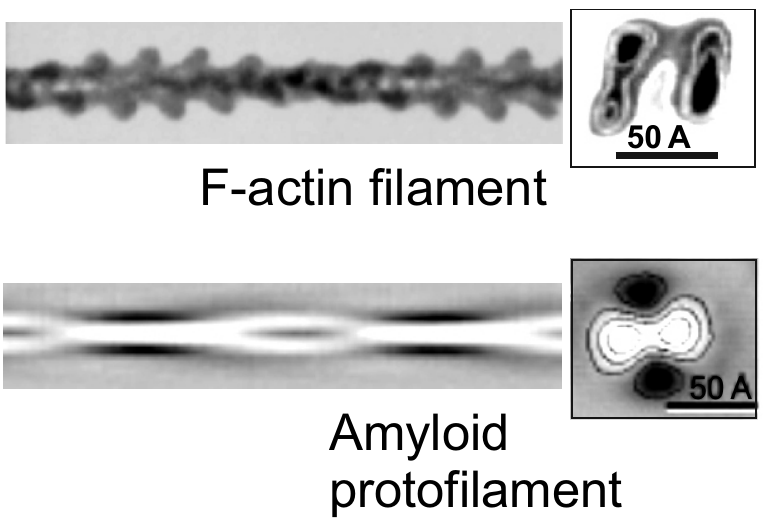}
\caption{Images reconstructed from cryo-EM experiments on actin filaments and
insulin amyloid protofilament, both showing a characteristic two-strand
structure of protein subunits bonded by physical interactions. In F-actin, the
growth occurs from the ATP-functionalized B-end and the dissociation from the
P-end where the F-actin subunits contain ADP. In the case of amyloids, the
growth and dissociation occur symmetrically from both ends.  }
\label{dual}
\end{figure}

To model a protein filament we use a coarse-grained model where the protein
monomers are treated as Brownian spherical particles of diameter $\sigma$
assembled into a two-stranded structure shown in fig. \ref{dual}. This
two-stranded topology (which includes the double-helix as one of its variants)
is one of the most commonly observed structures for both F-actin and amyloid
filaments. Each protein is interacting with two other proteins along the same
chain (longitudinal bonds), and with a third protein on the second chain
(transversal bond). The strength of the physical bonding is $\varepsilon_\|$
along the filament and $\varepsilon_\bot$ with the matching subunits in the
parallel strings. This is clearly a minimal model which however allows us to
capture the general features of two-strand filament breakup without the
complications that lie in the peculiar chemistry of different proteins. Hence,
it is hoped that the model predictions capture essential features that are
common to both actin filaments and amyloid fibrils.

\begin{figure}
\centering
\includegraphics[width=0.99\columnwidth]{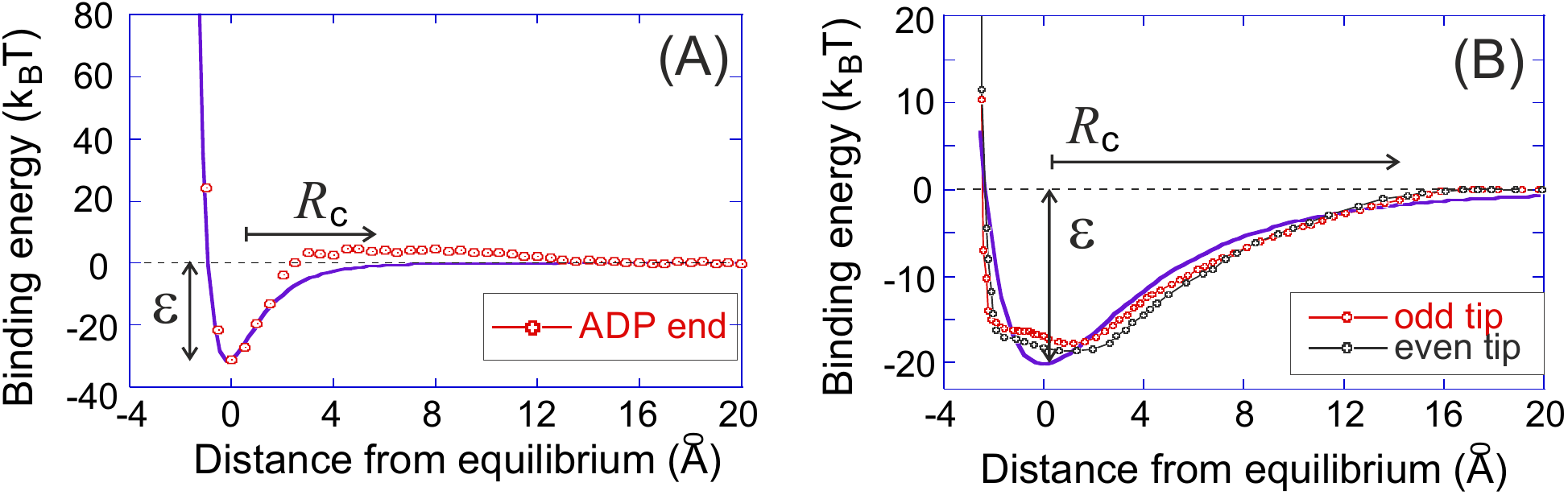}
\caption{(A) The molecular-dynamics calculation of the effective binding
potential of a single actin-ADP subunit and the remaining filament at the
P-end. The plot shows the two key parameters: the cutoff distance $R_c$ and the
potential depth $\varepsilon$, which is in good agreement with previous
experimental estimates \cite{9-Erickson}.  (B) A similar molecular-dynamics
calculation of the effective binding potential for a dimer of A$\beta$-amyloids
taken from Ref.\cite{25-Han}. In both graphs the data points are fitted with
the (12-6) Lennard-Jones potential, showing a good qualitative agreement
(although with expected differences at the long-range attractive tail given the
complex shape of the subunit): the depth of the binding well, typically on the
order of 20-30$k_\mathrm{B}T$.  }
\label{phenomena}
\end{figure}

\begin{figure}[t]
\centering
\includegraphics[width=0.5\columnwidth]{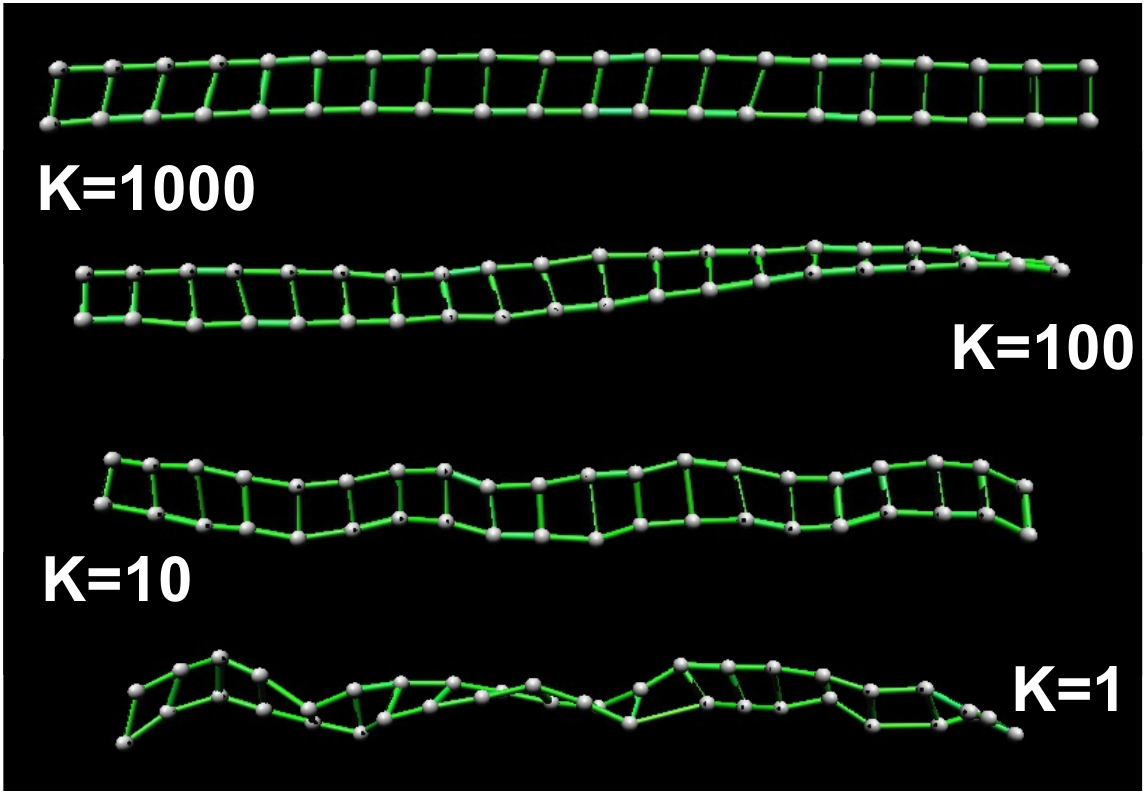}
\caption{An illustration of the role of bending rigidity: instantaneous shapes
of a coarse-grained two-stranded filament with increasing bending modulus $K$
(in units of $k_\mathrm{B}T$).  }
\label{stiff}
\end{figure}

Interactions between two proteins are modelled as the sum of a non-covalent
Lennard-Jones (LJ) interaction, and a bond-bending angular interaction. The LJ
potential describes the short-range steric repulsion between two proteins,
which co-exists with a central-force (London-van der Waals) attraction due to
hydrogen bonds holding two $\beta$-sheets together (prevalent in amyloids) – or
the hydrophobic interaction with a subunit perpendicular to the chain (dominant
in F-actin), see fig. \ref{phenomena}. The bond-bending term, instead,
originates from the geometrical constraint imposed by the $\beta$-sheet
connection or by other anisotropic steric interactions. In simple terms, when
two planar surfaces ($\beta$-sheets) are connected by several springs (hydrogen
bonds), any tangential displacement (orthogonal to the direction connecting the
two centers of mass of the two proteins) costs a finite amount of energy,
because of the rotational symmetry-breaking. The local bond-bending modulus
$K$, in units of $k_\mathrm{B}T$, is directly related to the persistence length
$l_p$ of the filament via the standard expression: $l_p=K \sigma
/k_\mathrm{B}T$. In fig. \ref{stiff} filaments with different values of bending
stiffness $K$ are shown, to illustrate the effect $K$ has on the overall
stiffness/flexibility of the filament.

To calibrate our coarse-grained potential on realistic filaments, we can
therefore use the experimentally known persistence length $l_p$ to guide our
choice of $K$ in the simulations. For biological filaments the persistence
length is long, though not infinite. While it can reach up to 1mm for
microtubules, it is typically about 18$\mu$m $\approx$3600 subunits for F-actin
\cite{21-Gittes} and 3$\mu$m $\approx$5000 subunits for
 amyloid-like insulin filaments \cite{22-Knowles}. These persistence lengths in
our simulations can be reproduced using large values of bond-bending stiffness:
$K \approx 10^4$ (in units of $k_\mathrm{B}T$). We have checked that our
results for the breakup rate and fragment distributions did not vary when $K$
is in the range $10^3$-$10^5$.

\subsection{Calibration of simulation parameters}
Next, we consider the parameters which control the LJ interaction, namely the
magnitude of binding energy $\varepsilon$ and the cut-off $R_c$.  To calibrate
these parameters, we first consider how they affect the breakup rates and
fragment distributions. In our simulations, both the end-dissociation of a
subunit and the filament fragmentation are directly observed, and their rates
recorded. Fragmentation can occur in two ways. Either two longitudinal bonds
break up which are the mirror-image of each other (same position along the
filament), thus leading to two fragments which both contain an even number of
subunits, or three (very seldom more) bonds break up, of which one is a
transversal bond, leading to two fragments both containing an odd number of
proteins. Clearly, the first breakup mode is energetically more favorable as
fewer bonds need be broken, two instead of three/more, and thus it occurs more
frequently. We declare a bond broken when the distance between the two subunits
exceeds the cut-off separation $R_c$, at which the attraction energy between
two proteins is set to zero.  Finally, we always keep the same constant ratio
between $\varepsilon_\| = \varepsilon$ for longitudinal bonds and
$\varepsilon_\bot = \varepsilon /2$ for transverse bonds. The chosen ratio
$\varepsilon_\bot / \varepsilon_\| = 0.5$ is somewhat arbitrary, but this value
has been deduced in the studies of amyloids \cite{23-Knowles}. For F-actin, the
situation is not much different, with a reported ratio $\varepsilon_\bot /
\varepsilon_\| = 0.67$ in a slightly different topology \cite{9-Erickson}
(as schematically depicted in Appendix A, Fig. \ref{si1}, and as discussed
extensively in the discrete model of Ref.~\cite{Kolomeisky_JCP}). This
similarity allows us to treat F-actin and amyloid within a common framework and
to interpret results for both systems.

We have performed simulations to span the $\varepsilon-R_c$ plane in a
computationally accessible region, by keeping $K=10^4$ fixed. The results are
reported in fig. \ref{stretch}. For all conditions investigated, the fragment
size distribution is strongly U-shaped, with the highest breakup rate occurring
for monomers at the ends of the filament, while the breakup rates in the inner
locations are mostly uniform and weakly dependent on the position along the
filament. We shall notice the characteristic even-odd behavior of the breakup
rate in the inner locations, discussed above.

\subsection{The qualitative role of the coarse-grained interaction potential}
\begin{figure}
\begin{center}
\includegraphics[width=0.95\columnwidth]{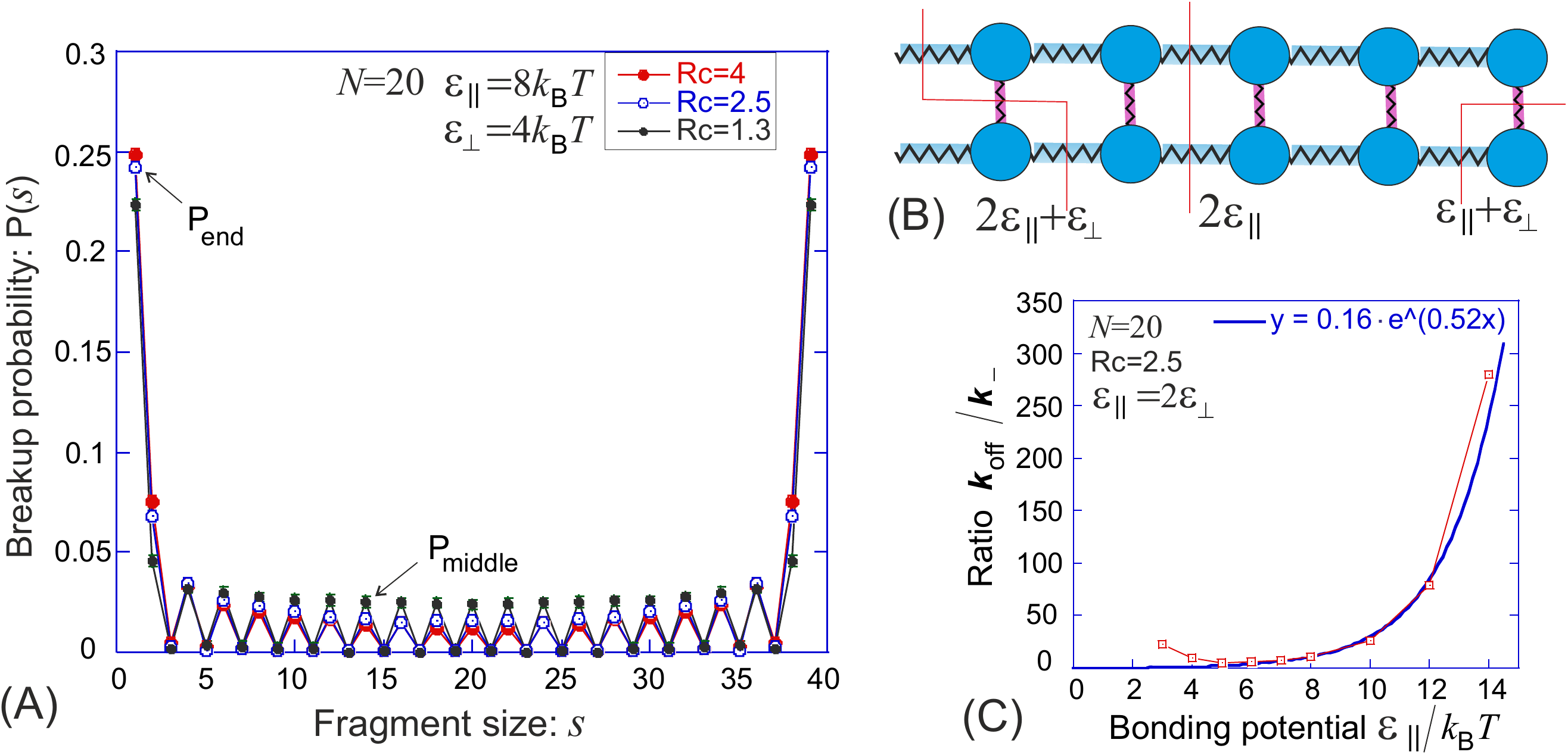}
\caption{(A) The result of our calculation of statistical probability of
thermal breakup of a two-stranded filament (parameters are labelled in the
plot, note that we take the transverse binding to be the half of the
longitudinal binding strength). The fragment size of 1 (or 39, for the
two-stranded filament of length N=20) represents the single subunit
dissociation; we can see a clear even-odd effect with the higher probability
for the filament to break clean across the middle rather than have a
complex-shape break. In (B) a sketch of this breakup mechanism is reported. The
three data sets in the plot (A) show that there is no strong dependence on the
cutoff distance $R_c$ of the potential. In (C) we plot the ratio of the
end-dissociation rate $k_\mathrm{off}$ to the fragmentation rate $k_{-}$, as a
function of strength of the physical bond $\varepsilon_\|$. At weak bonding,
the reaction rate deviates from the Kramers thermal-activation law, but this is
unlikely to be a relevant regime; at significant bonding strength the
end/middle rate ratio clearly follows an exponential: $\exp [0.52 \varepsilon /
k_\mathrm{B}T ]$ according to the fitting of our data.}
\label{stretch}
\end{center}
\end{figure}

The fragment size distribution would be flat, for stiff filaments, with equal
rates for end-dissociation and fragmentation, if the potential were symmetric
(e.g. harmonic or quartic). This fact can be explained qualitatively based on
the various contributions to the partition functions of fragments and their
dependence on fragment size, when local bending rigidity is active
\cite{26-Zaccone}. The much higher breakup probability at the ends is due to
the asymmetry of the LJ potential, which makes it much easier for the protein
at the end to escape from the bonding minimum in the outward directions where
the inflection point in the potential marks the upper-bound of restoring force.
This effect is captured by varying the cut-off $R_c$, because this parameter
controls the asymmetry of the LJ potential (keeping the curvature in the
minimum fixed). As shown in fig. \ref{stretch}(A), the breakup probability at
the end increases upon increasing $R_c$, and with it the asymmetry of the LJ.
This effect is negligible in the inner filament where the effective well
explored by a fluctuating subunit is more symmetric due to the
nearest-neighbors on both sides.

\subsection{Simulation results}
The probability of a breakup event is directly proportional to the rate
constant. We find that the ratio between the breakup rate at the filament end
and the rate of fragmentation in the inner part of the filament is a strongly
increasing function of $\varepsilon$, with an expected qualitative trend which
is exponential (Arrhenius), $k_\mathrm{off}/k_- \approx \exp [0.52 \varepsilon
/ k_\mathrm{B}T ]$, as illustrated in fig. \ref{stretch}(C). This result can be
rationalized using a simple argument based on the different connectivity at the
end and in the middle. According to the Kramers escape rate
theory~\cite{27-Kramers,28-Zaccone}, the end-subunit escapes from the bonding
minimum, with an exponential dependence on the total energy barrier,
$k_\mathrm{off} \approx \exp [-( \varepsilon_\| + \varepsilon_\bot) /
k_\mathrm{B}T ] = \exp [-1.5 \varepsilon / k_\mathrm{B}T ]$, since two bonds
(one longitudinal and one transverse) need be broken, see fig.
\ref{stretch}(B). Breakup in the middle, instead, involves the breaking of two
longitudinal bonds; breakup of three bonds (two longitudinal, one transverse),
or even more, is practically negligible because more cooperative and
energetically unfavorable motion is required. Therefore we have $k_- \approx
\exp [-2 \varepsilon_\| / k_\mathrm{B}T ] = \exp [-2 \varepsilon /
k_\mathrm{B}T ]$. Upon forming the ratio, we readily obtain $k_\mathrm{off}/k_-
= \exp [( \varepsilon_\| - \varepsilon_\bot) / k_\mathrm{B}T ] = \exp [0.5
\varepsilon / k_\mathrm{B}T ]$, in excellent agreement with the law found in
simulations (the small deviation from 0.5 is certainly due to a proportion of
rare complex-topology breaks). This result is very important because it
establishes that, due to the different connectivity of subunits in the
filament, the ratio between end-dissociation and fragmentation rates has to
increase exponentially with the protein-protein binding energy $\varepsilon$.

\subsection{Application to F-actin}
For F-actin, Erickson\cite{9-Erickson} has estimated the difference between the
energy barriers for fragmentation and end-dissociation: $\sim$10.7 kcal/mol =
17.9$k_\mathrm{B}T$, giving the prediction $k_\mathrm{off}/k_- \approx 6 \cdot
10^7$. This compares very well with the data assembled by Pollard and Cooper
\cite{19-Pollard}, who quote the range for $k_\mathrm{off} \approx$ 0.5-5
$\mathrm{s}^{-1}$ and $k_- \approx 10^{-8}$ $\mathrm{s}^{-1}$.
It should be noted that the fragment size distribution in the case of actin is,
in reality, not symmetric at the two ends in the fast-polymerization or
treadmilling limit, where the pointed end is made of ADP-actin subunits,
whereas the barbed end is formed by ATP-actin, for which depolymerization is
suppressed. The end-dissociation rates are $\sim$0.3 $\mathrm{s}^{-1}$ and
$\sim$2 $\mathrm{s}^{-1}$ for the P-end and for the B-end, respectively,
according to Pollard \cite{8-Pollard,19-Pollard}. However, the difference is
less than an order of magnitude and does not introduce any substantial
qualitative change in our picture. The rates are equal at both ends, and the
distribution symmetric, in the opposite limit of slow polymerization at low
monomer concentrations \cite{2-Alberts}.

Appendix A and Fig. \ref{si1} give more detail about F-actin filament
and its bond structure. Applying the same analysis here with Pollard's $2/3$
ratio for $\varepsilon_\bot/\varepsilon_\|$, we obtain $k_\mathrm{off} \approx
\exp [-( \varepsilon_\| + \varepsilon_\bot) / k_\mathrm{B}T ] = \exp
[-\frac{5}{3} \varepsilon / k_\mathrm{B}T ]$, and
$k_- \approx \exp [-(2 \varepsilon_\| + \varepsilon_\bot)/ k_\mathrm{B}T ]=\exp
[-\frac{8}{3} \varepsilon / k_\mathrm{B}T ]$.
Hence, it follows that $k_\mathrm{off}/k_- = \exp [ \varepsilon/
k_\mathrm{B}T]$, that is, an Arrhenius dependence on the longitudinal binding
energy. Comparing with the result for amyloids, fig. \ref{stretch}, the ratio
between dissociation rate at the end and breakup rate in the middle is much
larger for actin. This, for typical binding energy on the order of $20
k_\mathrm{B}T$ (conservative estimate) gives an additional factor of $\sim
e^{10}$ for actin $k_\mathrm{off}/k_- $ ratio, with respect to the case of
amyloid. This means that dissociation rate at the end over breakup in the
middle is $\sim e^{10}\approx 2.2\times10^{4}$ times larger for actin than for
amyloids. This estimate suggests that while breakup in the middle may play an
important role in amyloids, it can be safely neglected in the dynamics of
F-actin fibres.

In the remaining of this paper we focus on the amyloid system, for which the
`ladder' bond structure illustrated in fig. \ref{stretch}{B) holds, and for
which new experimental results are reported below.

\subsection{Application to amyloid-like fibrils and comparison with
experiments}
For amyloid fibrils, using typical values of binding energy for $\beta$-sheets
bonding in amyloid-like aggregates, which are on the order of 20-30$
k_\mathrm{B}T$,\cite{29-Dobson} our approach yields the prediction of
$k_\mathrm{off}/k_- \approx 10^5$.
In order to verify our prediction and the proposed molecular mechanism in
amyloids, we experimentally determined the end-dissociation rate for insulin
amyloid filaments exhibiting the typical two-stranded structure used in our
model calculations, and compared it with the fragmentation rate estimated
previously from kinetic fitting of total mass and length distribution: $k_-
\approx 10^{-9}-10^{-8}$ $s^{-1}$ at T=60${}^\circ$C \cite{16-Knowles,24-Smith}. In
this paper we obtain an estimate of $k_\mathrm{off}$ for insulin amyloid
fibrils through direct observation of monomer release into solution as a
function of time, as illustrated in fig. \ref{Grelax} and more details can be
found in Appendix B. To this effect, a sample of mature insulin amyloid fibrils
was first prepared at pH 2.0; we then isolated the fibrils by
ultracentrifugation, and the supernatant, containing free monomer, was removed
and replaced with a dilute aqueous solution of HCl, also at pH 2.0. In this
manner we depleted an insulin amyloid fibril sample free of monomer. Over time,
monomer dissociation from the fibril ends re-established the monomer-fibril
equilibrium in the supernatant. To probe for soluble insulin during this
process, we ultracentrifuged sample aliquots during a time course and
investigated the supernatant solutions by gel electrophoresis as shown in fig.
\ref{Grelax}(B). These end-dissociation experiments were carried out at
T=60${}^\circ$C, to match with existing literature data on $k_-$ , and also at
T=4${}^\circ$C to allow better time-resolution of slower kinetics and to
estimate the free energy barriers involved.

\begin{figure}
\centering
\includegraphics[width=0.75\columnwidth]{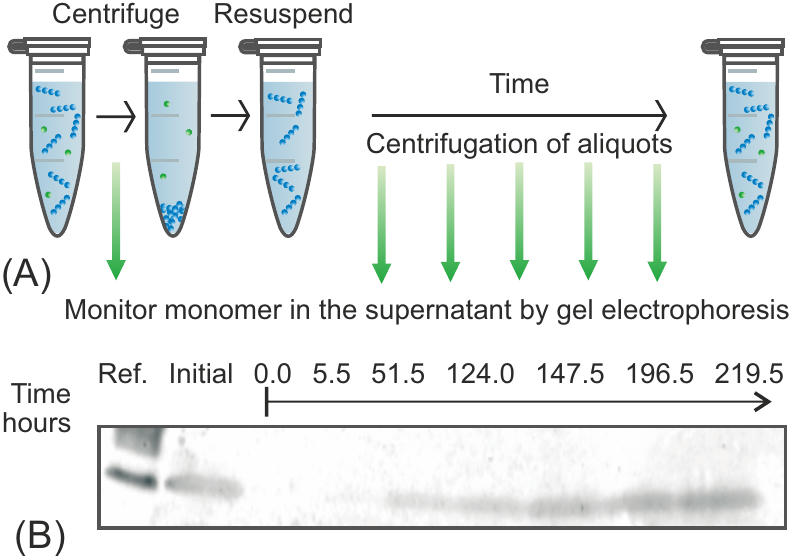}
\caption{Experimental determination of monomer detachment rates from insulin
amyloid fibrils. (A) A schematic showing how insulin fibrils were depleted of
monomer and subsequently incubated. (B) The establishment of an equilibrium
monomer population was monitored by SDS PAGE. Here a gel for a time course at
4${}^\circ$C is shown. From left to right, the gel contained: the reference
band from a protein standard; the monomer from the initial fibril sample
supernatant; the supernatants from the time points from 0 to 219.5 hours after
the removal of the initial monomer population.   }
\label{Grelax}
\end{figure}

Assuming that fibril fragmentation is much slower compared to depolymerization,
the concentration of free monomer as a function of time, $m(t)$, is given by
the equilibrium monomer concentration, $m_\mathrm{eq}$, the number
concentration of fibrils, $P$, and the dissociation rate constant
$k_\mathrm{off}$, as: $m(t) =  m_\mathrm{eq} [1-\exp (-2 P k_\mathrm{off}
t/m_\mathrm{eq})]$. We determined a value for the product $ P k_\mathrm{off}$
through a least squares fit to the data in fig. \ref{Grelax2}. To find
$m_\mathrm{eq}$ we calibrated the concentration at the plateau against a
dilution series of known concentration, see Methods below. This analysis gave a
value of  $ P k_\mathrm{off} = 1.3 \cdot 10^{-10}$ M/s at 60${}^\circ$C and
$1.2 \cdot 10^{-12}$ M/s at 4${}^\circ$C. Considerations based on the average
density, diameter and length of fibrils \cite{24-Smith} suggest that the
average number of monomers per fibril is ca. 5000 (8.2 monomers per nm). Given
the mass concentration of fibrils (2mM), this yields the concentration of
fibrils $P = 4 \cdot 10^{-7}$ M, then $k_\mathrm{off} = 3.3 \cdot 10^{-4}$ $s^{-1}$
at 60${}^\circ$C. The same analysis gives $k_\mathrm{off} = 3 \cdot 10^{-6}$
1/s at 4${}^\circ$C. Taking the value for the fragmentation rate of the same
insulin fibrils, as measured previously \cite{16-Knowles,24-Smith}, the ratio
becomes: $k_\mathrm{off}/k_- \approx 1.5 \cdot 10^5$ at 60${}^\circ$C.

\begin{figure}
\centering
\includegraphics[width=0.75\columnwidth]{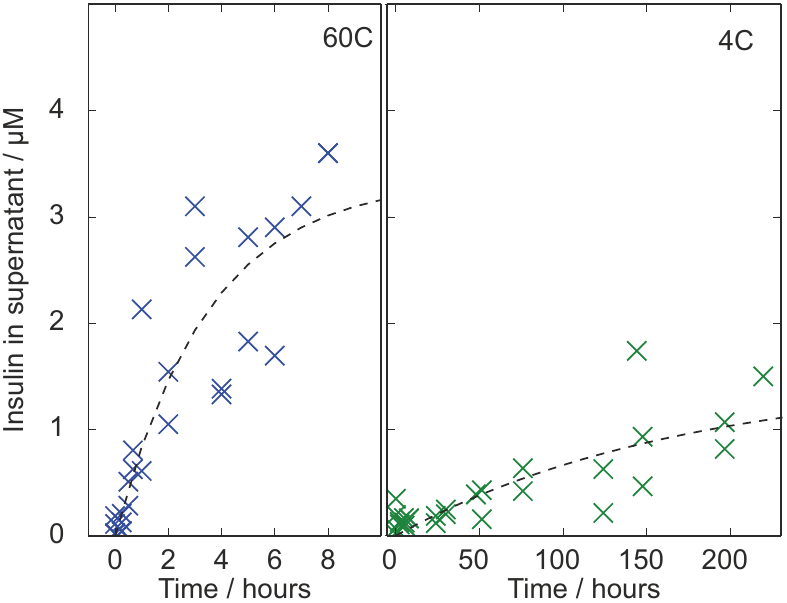}
\caption{ The measured concentration of soluble insulin against time, and the
fit to the data, dashed line, at T=60${}^\circ$C and 4${}^\circ$C.   }
\label{Grelax2}
\end{figure}

\section{Conclusions}
Using the theoretically predicted (exponential) result in fig. \ref{stretch}(C)
(and the value of T=60${}^\circ$C), we obtain the strength of the longitudinal physical bond
between insulin subunits in the amyloid filament: $\varepsilon_\| \approx$ 13
kcal/mol = 22$k_\mathrm{B}T$. This is very close to our own
ab-initio simulations of two-stranded F-actin in fig. \ref{phenomena}(A), the results of
molecular-dynamics work on amyloids \cite{25-Han} reproduced in fig. \ref{phenomena}(B),
and the common-sense expectation for a sequence of hydrogen bonds between two
$\beta$-sheets. It also confirms our a priori assumption that the transverse
bonding (mainly due to hydrophobic interactions) is approximately half in
strength of $\varepsilon_\|$.

The availability of values for $k_\mathrm{off}$ at two different temperatures
allows us to estimate the enthalpic barrier $\Delta H_2^\ddagger$ for the
dissociation process, since $\ln [k_\mathrm{off}(1)/k_\mathrm{off}(2)] = \Delta
H_2^\ddagger \cdot (1/k_\mathrm{B}T_2 - 1/k_\mathrm{B}T_1)$. This yields
$\Delta H_2^\ddagger \approx$ 16 kcal/mol, very close to the value of enthalpic
barrier $\Delta H_1^\ddagger$ for the forward process measured previously
\cite{30-Buell}.

\begin{figure}
\centering
\includegraphics[width=0.8\columnwidth]{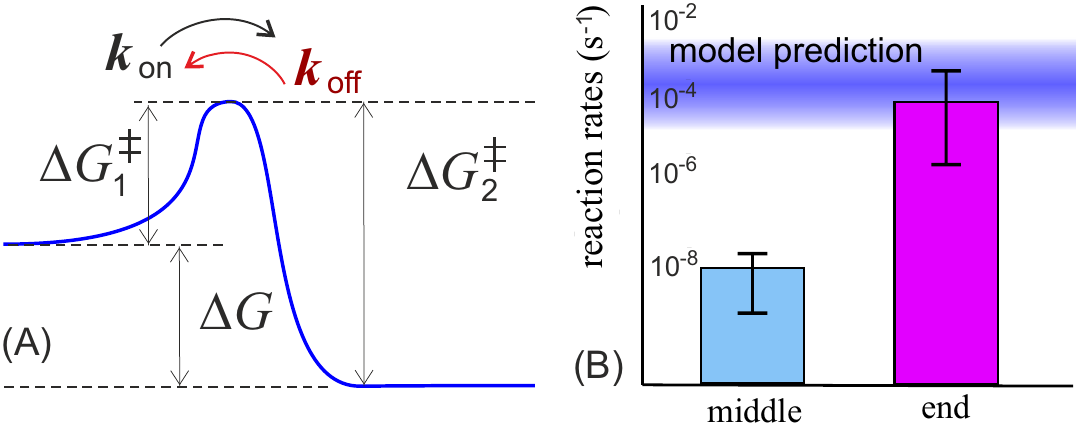}
\caption{(A) A scheme of free energy levels and the barriers for aggregation
and end-dissociation, $k_\mathrm{on}$ and $_\mathrm{koff}$, respectively; the
difference $\Delta G$ is essentially the bonding strength $\varepsilon$. It
should be noted that in our simplified coarse-grained model we neglect
rotational degrees of freedom (since our sub-units are perfect spheres), which
means that some entropic contribution to the free energy is necessarily
reapportioned to account for this - quantitatively precise estimates of free
energy barriers can only be extracted from ab-initio simulations. (B) A
pictorial comparison of the experimental data on breakup rates in insulin
amyloid fibrils, from Ref.24 and this work.  }
\label{answ}
\end{figure}

In summary, we have shown that thermal breakup in a minimal model of typically
two-stranded biomolecular filaments can be understood in a general way and
depends exponentially on: 1) the topology of connectivity, 2) the difference in
bonding energy in the longitudinal and transverse direction, and 3) the nature
and asymmetry of the protein-protein interaction. All these effects strongly
favour end-dissociation (detachment of a single protein from the filament ends)
over the fragmentation of filament into two large fragments. In particular, for
the most typical two-stranded structure observed in both actin and amyloid
filaments, we establish the general law $k_\mathrm{off}/k_- = \exp [(
\varepsilon_\| - \varepsilon_\bot) / k_\mathrm{B}T ]$ for amyloid fibrils, and
a similar estimate for F-actin fibres gives $k_\mathrm{off}/k_- = \exp [
\varepsilon/ k_\mathrm{B}T]$. This important parameter is influenced by the
nature of the protein-protein interaction (whether hydrophobicity- or hydrogen
bond-controlled). With realistic values of binding energy from $\beta$-sheet
bonding, this ratio reaches values on the order of $10^4$-$10^5$, an
order-of-magnitude result which we were able to confirm experimentally on the
example of insulin amyloid-like filaments, while the value for this ratio is
comparatively much larger, in the order of $10^8$-$10^9$, for F-actin. These
findings serve as the basis for improving the numerical description of protein
aggregation phenomena within a common quantitative framework and, possibly, in
future applications, for the development of pharmacologically-controlled
cleavage of protein aggregates in vivo.

\section*{Appendices}

\subsection{MD simulation of subunit binding in F-actin}

F-actin filament structure, as schematically depicted above in Fig.
\ref{si1}, was adopted from the extensive work of Voth et al.
\cite{Chu-Voth,Voth2011}, who have used a periodically repeating 13-monomer
segment of F-actin with subunit structures taken from
Protein Data Bank (PDB) structures 1J6Z (G-like; \cite{Otterbein}), 2ZWH
~\cite{Oda}, and 3MFP
~\cite{Fujii} and equilibrated it in waters with ADP as the bound nucleotide
and Mg2+ at the
high-affinity cationic bind site. Our starting point for the investigation of
an effective binding potential was
the PDB file of this 13-subunit long actin filament.

Our approach has been to fix all atom positions of the 12 subunits (A2-A13) and
simulate moving of the
center of mass of the terminal (A1) end-subunit along the line parallel to the
filament axis. The energy change in this move is the potential reported in fig.
\ref{phenomena}(A). We first moved
the whole A1 subunit with its atoms frozen with respect to its center of mass,
and then equilibrated the
ADP-bound actin molecule while keeping its center of mass fixed in its new
position. The movement step
of 0.5$\AA$ was found sufficient to give adequate resolution of the energy
function, see fig. \ref{phenomena}(A). The simulations were
performed using the CHARM22 force field with CMAP correction and modified
parameters for methylated
histidine \cite{31-Voth} and the NAMD simulation code ~\cite{NAMD}. Both
electrostatic and LJ potentials were truncated at a cutoff distance of 12$\AA$;
in fact, we discovered that CHARM22 parameters are optimized for 12$\AA$ and
larger cutoffs lead to distortions. Finally, the water (TIP3P model) was used
to solvate the proteins.

\begin{figure}[b]
\centering
\includegraphics[width=0.8\columnwidth]{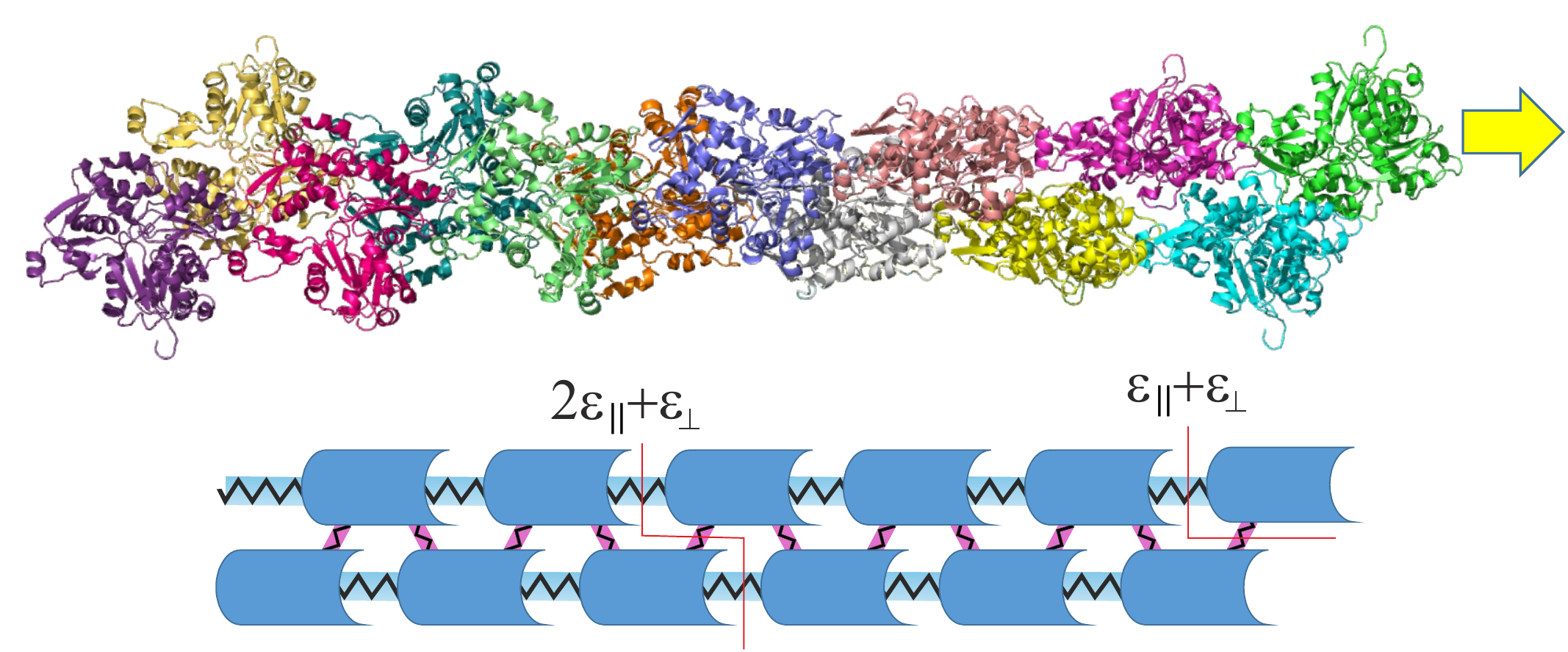}
\caption{Two-stranded structure of 13-subunits long ADP-actin filament used in
our simulations (the image is generated with PyMOL v. 1.7.7.2; the actin is
shown in cartoon representation with each subunit is colored differently). The
arrow shows the movement of the last subunit used to generate the energy
function.  The schematic of F-actin, indicating the bond-breaking energies,
differs from the `ladder' representation of amyloid filament in Fig.
\ref{stretch}(B).  }
\label{si1}
\end{figure}

\begin{figure}
\centering
\includegraphics[width=0.8\columnwidth]{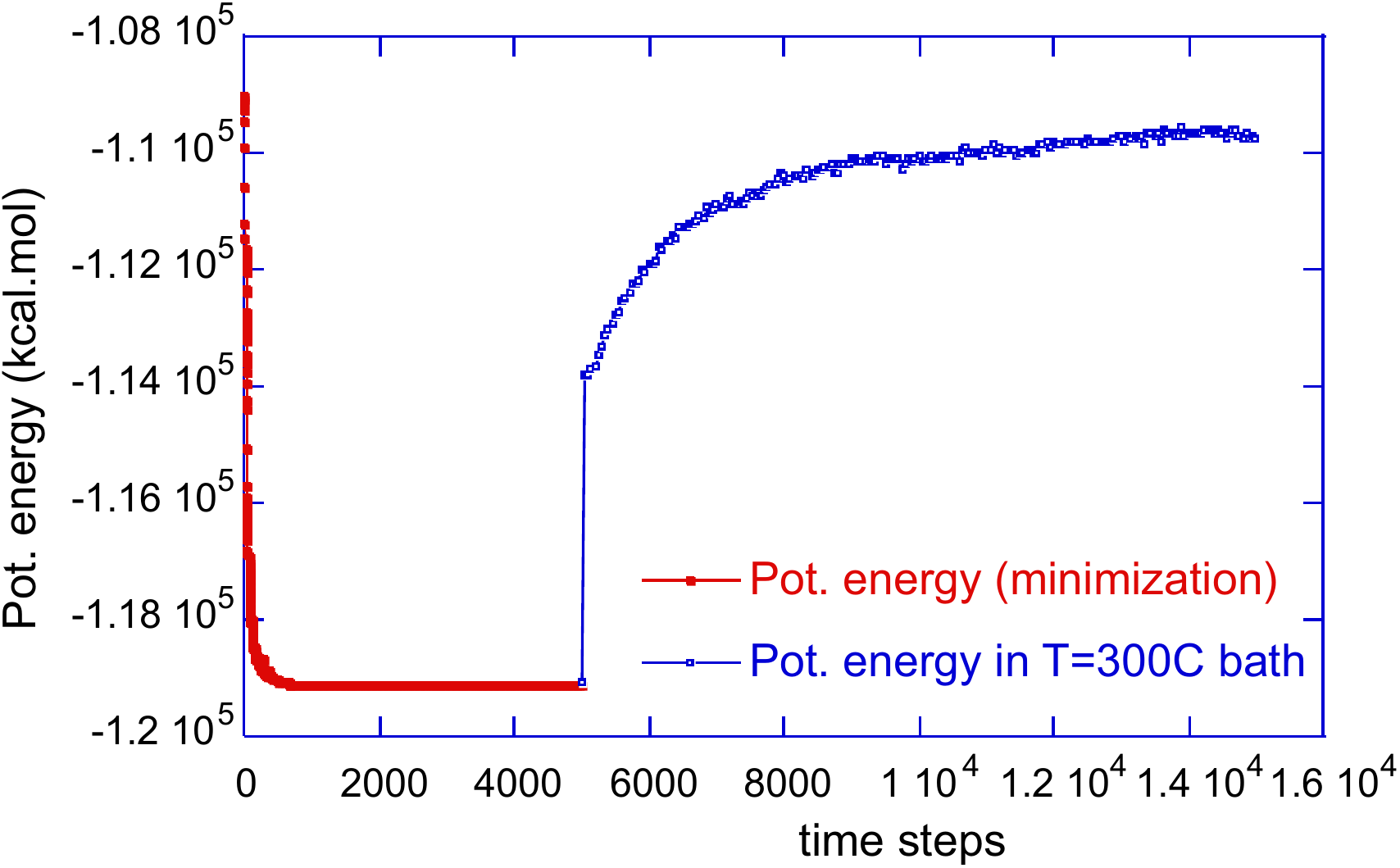}
\caption{The illustration of NAMD simulation output for one point (the distance
of 1.5$\AA$ from equilibrium for the end actin subunit being moved away from
the filament). The first part of the plot shows the rapid and efficient energy
minimization at T=0, followed by relatively slow equilibration in the heat bath
at 300K. }
\label{si2}
\end{figure}

For each simulation point (each position of the A1 center of mass) the
procedure involved the energy minimization for 5000 steps, after which the
system was heated to 300K and equilibrated for further 10000 steps. The
illustration is given in fig. \ref{si2}, for the point 1.5$\AA$ away from the
equilibrium along the stretching axis; it appears convincing that a
reasonable equilibration has taken place. After the whole sequence of
simulations was completed (spanning the distance of A1 center of mass of
-2$\AA$ to 20$\AA$, with zero defined as the equilibrium position) we have
taken the value of system energy at the maximum separation as zero. Then the
‘binding energy’ reported in fig. \ref{phenomena}(A)  is the change with
respect to that value: we recognize the deep attractive potential well and the
steep repulsive rise of energy on compressing the distance -- the
characteristic features of the LJ potential used in the subsequent
coarse-grained simulation.

\subsection{Coarse-grained simulation of bond breakup}

In our main computer simulation, we consider a coarse-grained model of protein
filaments as 1D chains of
Brownian particles where every particle is bonded to each of its two neighbors
via the truncated Lennard-
Jones potential, see fig. \ref{si3}(A), where the distance $r$ between the two
bonded particles is scaled by the parameter $\sigma$: is the hard-core diameter
of the
protein. We set the LJ potential to maintain a well depth equal to
$\varepsilon$ , independently of the chosen cutoff radius $R_c$.  When the
two-stranded filament is considered, the transverse bonds have the potential
depth  $\varepsilon / 2$ with the same cutoff.

We also include in our analysis the local bending energy. This effect is
reflected in the finite energy one
has to spend in order to bend the inter-protein bond, or equivalently, to
change the angle between two
adjacent bonds. Aiming to describe reasonably stiff filaments, we use the
bending energy in the form $\frac{1}{2} K
\theta^2$ , where $\theta_i$ is the angle between the directions of bonds from
the particle $i$ to the preceding $(i-1)$ and
the subsequent $(i+1)$ subunits. Figure \ref{si3}(B) illustrates the way this
effect is implemented by imposing
pairs of equal and opposite forces on the joining bonds, providing a net torque
on the junction. It is the same
algorithm as used in, e.g. LAMMPS `angle-harmonic' system \cite{LAMMPS}. The
bending modulus $K$, in units of $k_BT$,
is directly related to the persistence length of the filament via the standard
expression $l_p = K\sigma/k_BT$, where $\sigma$
is the ‘particle size’ in the LJ potential above.

\begin{figure} 
\centering
\includegraphics[width=0.7\columnwidth]{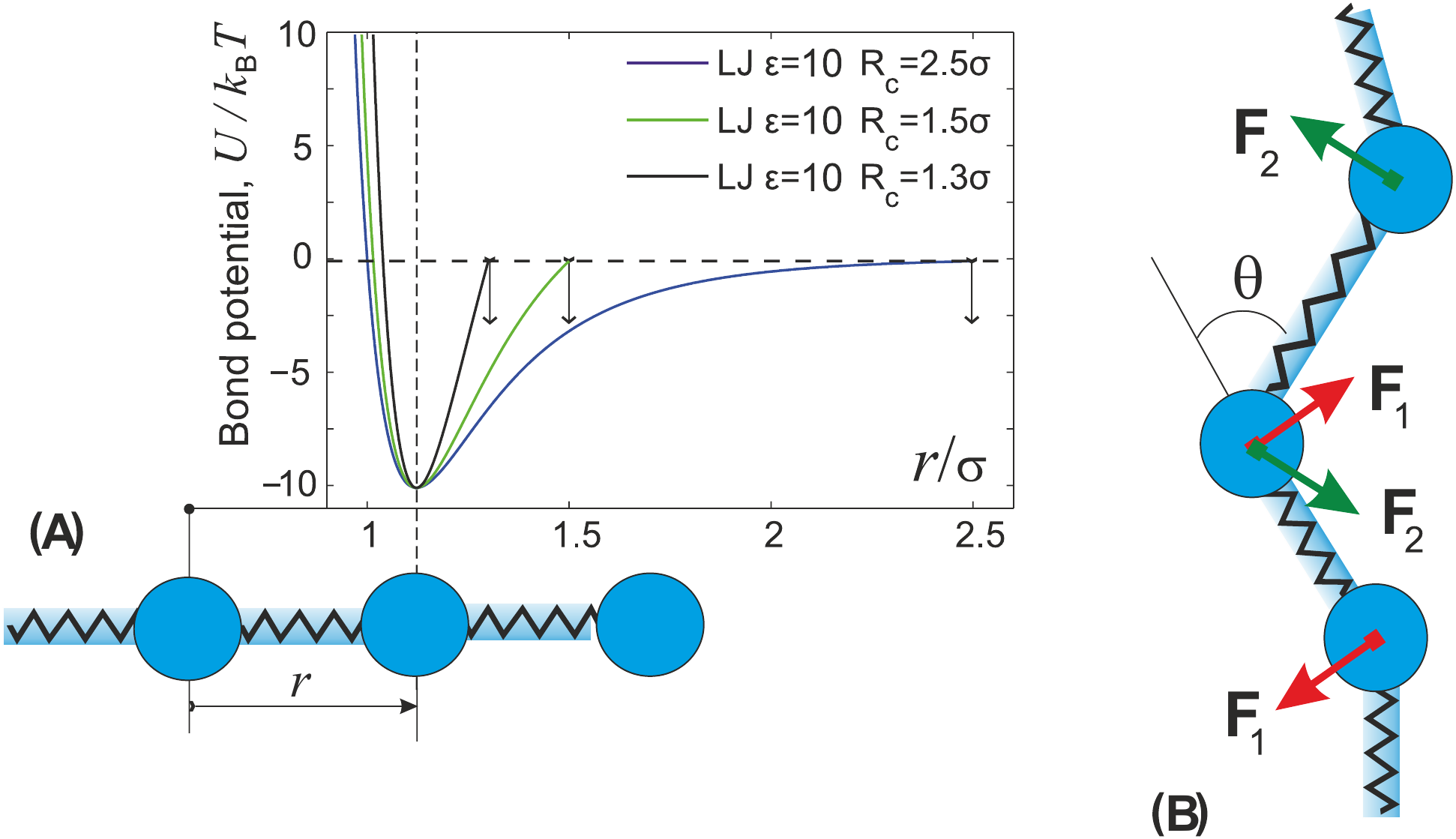}
\caption{(A) An illustration of the role of truncation of the LJ potential at
the cutoff distance $R_c$,
while maintaining the fixed depth of the potential well $\varepsilon$. At
larger $R_c$ the anharmonicity of the long-range
attractive potential becomes more prominent (crucially, affecting the
inflection point and the region of
unstable concave potential shape). (B) The bond-bending penalty is implemented
by forming two
pairs of equal and opposite forces (couples) on the adjacent bonds to affect
the bending angle $\theta$. }
\label{si3}
\end{figure}

The dynamics is governed by the overdamped Langevin equation, which is
discretized in the standard way; for each particle:
\begin{equation}
\bm{r}(t+\Delta t)= \bm{r}(t) - \frac{\Delta t}{\gamma} \nabla U(\{\bm(r)\}) +
\Gamma \sqrt{ \frac{2\Delta t}{\gamma}} \nonumber
\end{equation}
where $\bm{r}$ is the 3n-dimensional vector containing the positions of all
molecules, $\gamma$ is the friction coefficient
for one particle moving in the medium, and $\Gamma$ is the amplitude of
Gaussian stochastic force, defined according to the fluctuation-dissipation
theorem. We used the $\Delta t = 5 \cdot 10^{-5}$ in MD units, and $\gamma =1$
since the friction incorporated into the value of $\Gamma$. The equation was
integrated with an explicit Euler method.

Each run is initialized with interparticle distance $|r_i - r_{i-1} |= 2^{1/6}
\sigma$, corresponding to the minimum of the LJ
interaction potentials, and terminated when any of the bonds reaches the cutoff
$R_c$ . The location of the
rupture was recorded. To generate the adequate statistics, $N$ independent runs
were performed and the breakup probability calculated as $P_i= N_i / N$, where
$N_i$
was the number of recorded breakup events for the bond $i \in (1...19)$. We
used $N =10^6$ for the harmonic potential, $N = 5 \cdot 10^4$ for LJ potentials
with $\varepsilon \leq 4k_BT$ and $N = 3 \cdot 10^4$
for LJ potentials with $\varepsilon > 4k_BT$. Since runs were independent
from each other, the $N_i$ are binomially (Bernoulli) distributed and the error
bars were estimated as
$3\sqrt{ P_i (1-P_i)/N}$. For the case of insulin,
which has a diameter of about 2nm and $ D \approx  2\cdot 10^{-10} \mathrm{m}^2
/ s$,
the characteristic diffusion time is estimated as $\tau \approx$ 7ns. Our fixed
numerical time step is then
$\Delta t = (5 \cdot 10^{-5} ) \tau \approx 0.5$ps.

The simulation application was written in C++ and took advantage of multi-core
and multi-processor
capabilities of the executing hardware. In the scenario where the LJ potential
was acting in a repulsive
manner on two particles, the simulation code was designed to handle the
potentially unbounded resulting
force. It would theoretically be possible within a fixed time step to find a
pathological case where the force was too great and this would distort the
results by creating an artificial ejection leading to a breakup where otherwise
there would not have been one. To handle this an upper limit was selected on
the force and if a calculated component was equal or greater than this limit
the simulation would reset to the beginning of that time-step and re-execute it
in 5 smaller time-steps. This feature was logged and it was determined that
less than 0.25\% of total execution time was spent in this specific case.
Brownian motion components involved the use of pseudo-random number generation.
The number generation was implemented using the Mersenne Twister algorithm
\cite{Mersenne} and a normal distribution parameterised by the mean and
standard deviation was applied to the results. To ensure that no simulations
were executed using a seed for the number generator that matched a prior
execution run, seeds were taken to be fully 64-bit as opposed to the more
customary 32-bit approach, and based on a unique temporal component selected as
execution start time in compute cycles on the machine. This reduces the
probability of matching/duplicate seeds to effectively zero.
Finally the simulation itself performed multiple simulations concurrently. Each
simulation was executed on a single thread with pre-emption for more effective
resource scheduling, and the total number of simulations being executed in
parallel was derived from the absolute total number of available hardware cores
(accounting for extra thread capability from Hyper-Threading technology). This
was done to allow more simulations to be run to satisfy the need for a
statistically significant number of overall executions. In order to maximize
performance simulation code was written in a lock-less fashion and the
prevention of concurrency issues was accomplished through the use of
Interlocked (or more colloquially ‘atomic’) operations on key synchronisation
components.

\subsection{Insulin amyloid dissociation measurements}

Insulin fibrils were prepared using 0.5 mL 2 mM bovine insulin, 20 mM NaCl, HCl
pH 2. The solution was filtered using a syringe driven filter (0.22 $\mu$m pore
size, Millex), 1\% v/v preformed fibrils were added to seed the reaction, the
sample was then incubated at 65$^\circ$C overnight. 200 $\mu$L of the mature
fibril sample was then centrifuged in a Beckman ultracentrifuge at 90,000 rpm
for 15 minutes at 4$^\circ$C. These settings were used for all
ultracentrifugation steps. The supernatant was kept as a reference for the
monomer concentration. The pellet from 200 $\mu$L of the mature fibril solution
was re-suspended in 2 mL 10 mM HCl and centrifuged in 200 $\mu$L aliquots. The
supernatants from these were removed and replaced with 10 mM HCl, marking the
beginning of the time course. The samples were incubated at either 4$^\circ$C
or 60$^\circ$C to allow monomer dissociation. At each time point an aliquot was
ultracentrifuged and the supernatant probed for soluble insulin as shown in
fig. \ref{Grelax}.

The time courses were repeated twice, two gels were run for each time course.
For each temperature, the data from four gels has been combined in fig.
\ref{Grelax}(C). The SDS PAGE gels (NuPAGE 4-12\% Bis-Tris, Life Technologies)
were run in MES buffer at 200 V for 25 min using standard procedures, and
protein bands were stained using the Silverquest kit (Life Technologies)
according to the manufacturer instructions. On an SDS gel, insulin migrates as
the two component peptides, the protein standard ladder seen in fig.
\ref{Grelax}(B) is therefore the 3 kDa band (insulin B chain) from the SeeBlue
Plus 2 prestained standard (Life Technologies). The protein standard was also
used as a band intensity reference between gels.

In order to evaluate $m_\mathrm{eq}$, gel electrophoresis was performed for the
time points at the plateau of the time courses and a dilution series of insulin
solutions. The results of a calibration gel for the time course at
4$^\circ$C are shown in fig. \ref{si4}. We obtained a calibration curve from a
linear fit to the integral of the band intensity for the samples of known
concentration. The equilibrium monomer concentration at 4$^\circ$C was then
found to be 1.5 $\mu$M and the one at 60$^\circ$C to be 3.3 $\mu$M. The band
intensities were measured with the gel image analysis tool in
Ref.~\cite{Schneider}. The sum of the pixel intensities for each band was
recorded as the band integral. For each gel, a line of best fit was determined
for the signal intensity against the insulin concentration for the solutions
of known concentration (purple points in fig. \ref{si4}). This linear
relationship was then used to determine $m_\mathrm{eq}$ from the band
intensity at the plateau of a time course (green point in fig. \ref{si4}).

\begin{figure} [t]
\centering
\includegraphics[width=0.6\columnwidth]{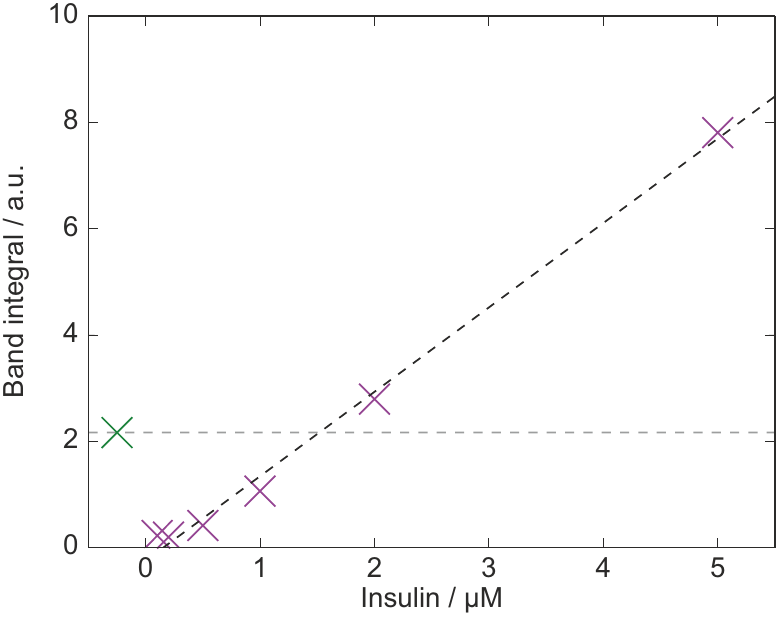}
\caption{Calibration of the equilibrium monomer concentration. A linear fit to
the integral of the band intensity for insulin solutions of known
concentrations, purple. The time point at 219.5 hours from the experiments at
4$^\circ$C was run on the same gel, green, we used the band integral of this
fraction to determine $m_\mathrm{eq}$. }
\label{si4}
\end{figure}

\subsection*{Acknowledgements}
This research was supported by the ERC, EPSRC, BBSRC, and the Newman
Foundation. Simulations were performed using the Darwin supercomputer of the
University of Cambridge HPC Service (http://www.hpc.cam.ac.uk/), provided by
Dell Inc. using Strategic Research Infrastructure funding from the Higher
Education Funding Council for England. We are grateful to Dr. Marissa Saunders
\cite{31-Voth} for providing a PDB file of the equilibrated actin filament
structure used to generate Fig. 1.

\end{document}